\pdfoutput=1
\documentclass[a4paper]{jpconf}
\usepackage{graphicx}
\usepackage{amsmath}
\usepackage{amssymb}
\usepackage{latexsym}
\usepackage{wrapfig}
\begin{document}

\bibliographystyle{iopart-num}

\graphicspath{{images/}}
\DeclareGraphicsExtensions{.pdf,.png}
\pdfimageresolution 110

\setlength{\unitlength}{1mm}

\clubpenalty = 10000
\widowpenalty = 10000

\title{Detector and Event Visualization \\ with SketchUp at the CMS Experiment}

\author{Tai Sakuma$^1$ and Thomas McCauley$^{2,}$\footnote[3]{Present address: University of Notre Dame, Notre Dame IN 46556, USA}}

\address{$^1$ Texas A\&M University, College Station TX 77843, USA}

\address{$^2$ Fermi National Accelerator Laboratory, Batavia IL 60510, USA}

\ead{sakuma@fnal.gov}

\begin{abstract}
We have created 3D models of the CMS detector and particle collision
events in SketchUp, a 3D modelling program. SketchUp provides a Ruby
API which we use to interface with the CMS Detector Description to
create 3D models of the CMS detector. With the Ruby API, we also have
created an interface to the JSON-based event format used for the iSpy
event display to create 3D models of CMS events. These models have
many applications related to 3D representation of the CMS detector and
events. Figures produced based on these models were used in conference
presentations, journal publications, technical design reports for the
detector upgrades, art projects, outreach programs, and other
presentations.
\end{abstract}


\vspace*{-0.4cm}

\section{Introduction}
The \textit{Compact Muon Solenoid} (CMS) experiment~\cite{CMS} is one
of two general-purpose experiments at the \textit{Large Hadron
Collider} (LHC) at CERN. From the data collected during LHC Run-I,
which ended in early 2013, the CMS collaboration has published more
than 250 papers describing searches for supersymmetry and exotic
phenomena, measurements of QCD, electroweak, top, bottom, forward, and
heavy-ion physics, as well as the discovery of the Higgs
boson~\cite{cms-higgs}.

Event displays are valuable tools that find many uses in collider
experiments. These uses include validation of detector geometries,
development of event reconstruction algorithms, visual inspection of
reconstructed events, and also production of high-quality visual
images for public presentations. One approach to develop an event
display is to build ``from scratch'' by using a 3D graphic library,
such as OpenGL, and a graphical user interface (GUI) toolkit. In the
CMS experiment, \textit{Fireworks}~\cite{fireworks},
\textit{FROG}~\cite{frog}, and
\textit{iSpy}~\cite{ispy-chep2012,ispy-website}, all of which are
actively used, were developed in this approach.

Here, we took an alternative approach; we created 3D models of the CMS
detector and events in an already-existing 3D modelling application,
widely used by architects, mechanical engineers, graphic designers,
and other professionals: \textit{SketchUp}~\cite{sketchup}. This
approach allows us to use many attractive features of SketchUp: it has
a highly intuitive user interface and precise dimensions; it can
export and import 2D images in various \textit{raster} and
\textit{vector} formats; it can exchange 3D models with other
applications in several common formats, e.g., 3DS and COLLADA; it can
apply visual effects on models, such as shadows, fog, and different
rendering styles.

The CMS geometry and event data are converted into SketchUp by using
the SketchUp Ruby API~\cite{SketchUpRubyAPI}. Our collection of Ruby
scripts are available on the GitHub repository
SketchUpCMS~\cite{sketchup-github}. The following sections describe
the rendering of the CMS detector geometry and events.

\section{Detector Geometry Rendering}

\begin{wrapfigure}[16]{r}{7cm}
\centering
\hspace*{0.2cm}\includegraphics[width=7cm]{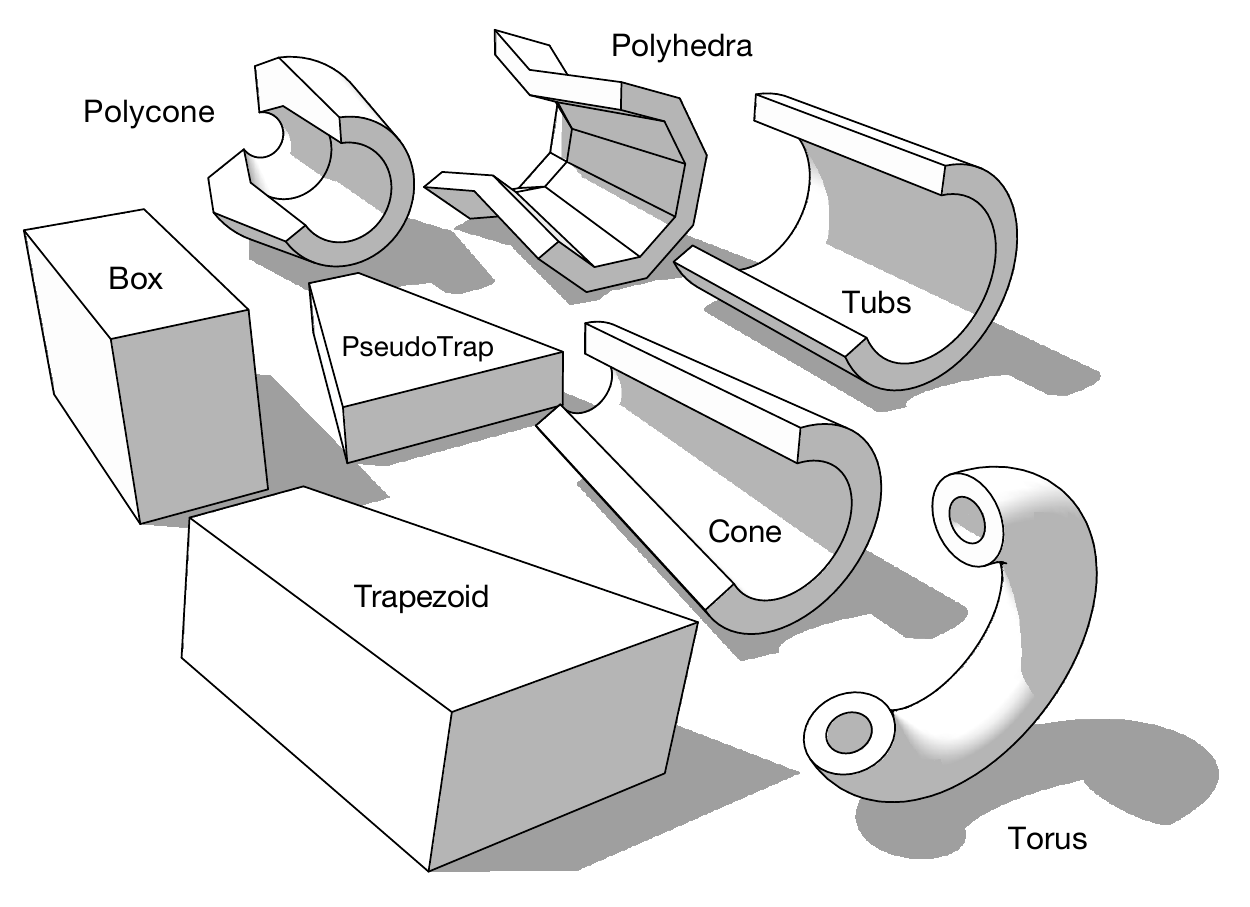}
\caption{\small Shapes of implemented solids} \label{solids}
\end{wrapfigure}

\begin{figure*}[!b]
\centering \includegraphics[width=14cm]{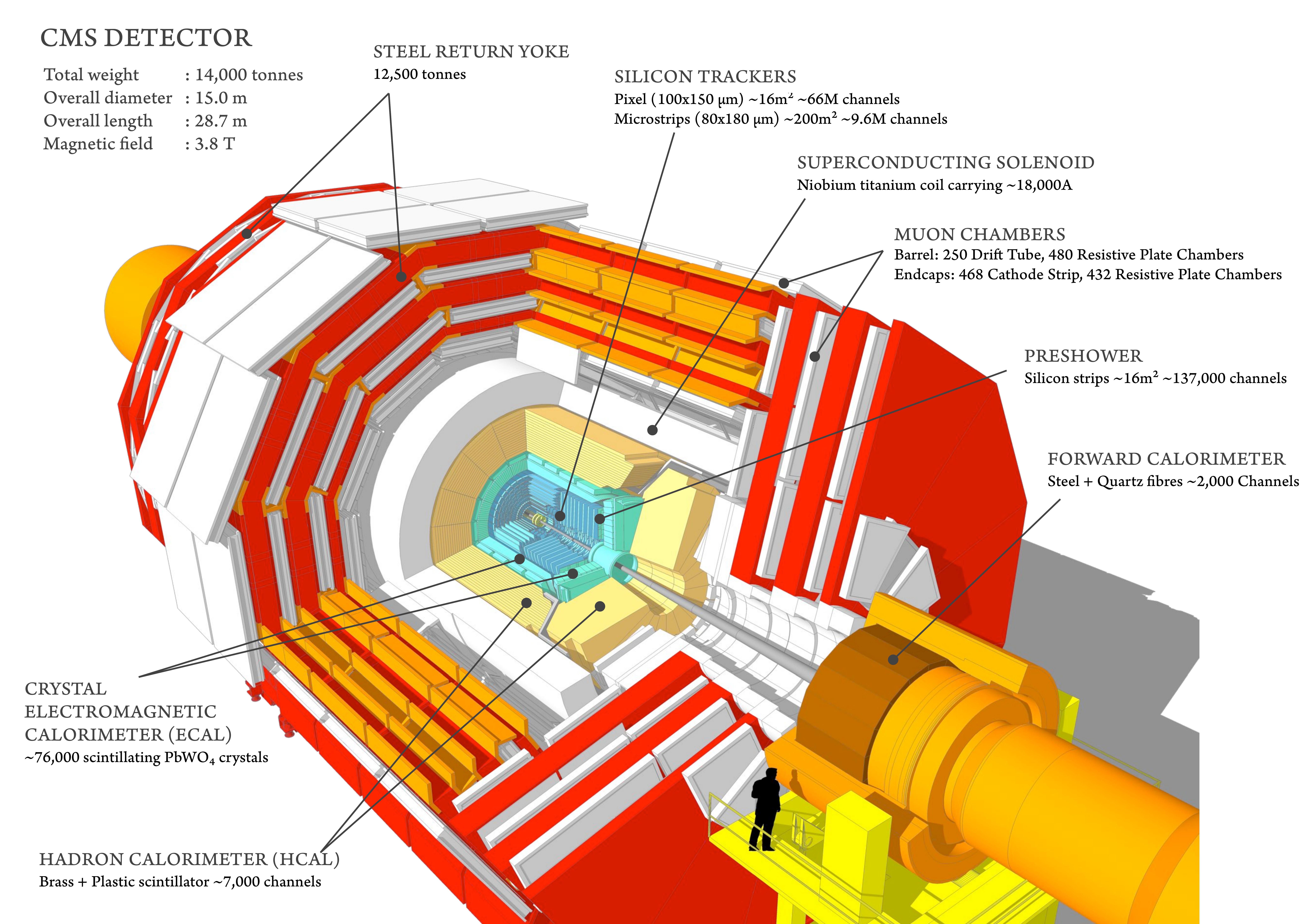} \caption{A
cutaway view of the CMS detector.} \label{fig:cms-image}
\end{figure*}

We obtain the detector geometry from the \textit{CMS Detector
Description}~\cite{cms-ddl}. The CMS Detector Description, written in
XML, is the master source of the CMS detector geometry used in the
\textit{CMS event reconstruction} and the \textit{CMS detector
simulation}. It describes the CMS detector as a \textit{directed
tree}. Each \textit{vertex} of the tree corresponds to a
\textit{component} with a size, shape, material, and density. Each
\textit{edge} connects from a component to its subcomponent; it
specifies the position and angle of the subcomponent within the
component.

The Ruby scripts take the following steps to render the geometry.
First, the Ruby scripts parse the XML files and recognize the directed
tree of the detector geometry. Second, they build each component as a
\textit{solid} with the given size and shape. Figure~\ref{solids}
shows the shapes and the names of implemented solids. Then, for each
edge from the \textit{leaves} of the tree to the \textit{root}, the
scripts place the \textit{tail} component in the \textit{head}
component as a subcomponent at the given position and angle.

Figure~\ref{fig:cms-image} is one of the CMS detector cutaway images
often used in public presentations. The 3D model in this figure was
rendered in SketchUp. This figure was used in the CMS official
website~\cite{cmsweb-detector}, posters created for CERN Open Days
\cite{web-cernopendays}, and the \textit{Higgs boson discovery
summary} published in \textit{Science}~\cite{higgs-science}.

\begin{figure*}[!t]
\centering \includegraphics[width=9cm]{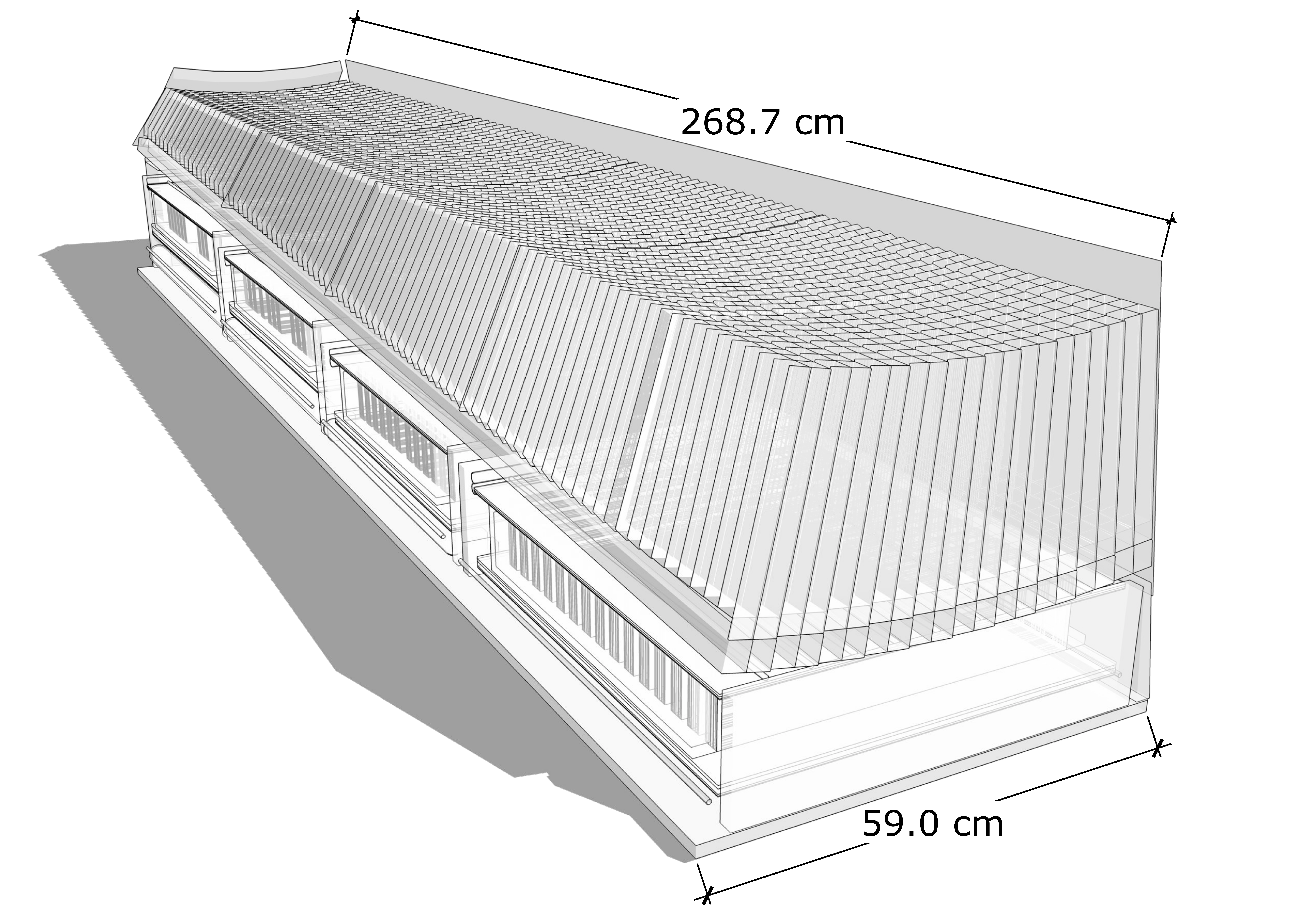}
\caption{\small A module of the Barrel Electromagnetic (EB)
Calorimeter} \label{fig:eb-module}
\end{figure*}

\begin{figure*}[!b]
\centering
\fbox{\includegraphics[width=14cm]{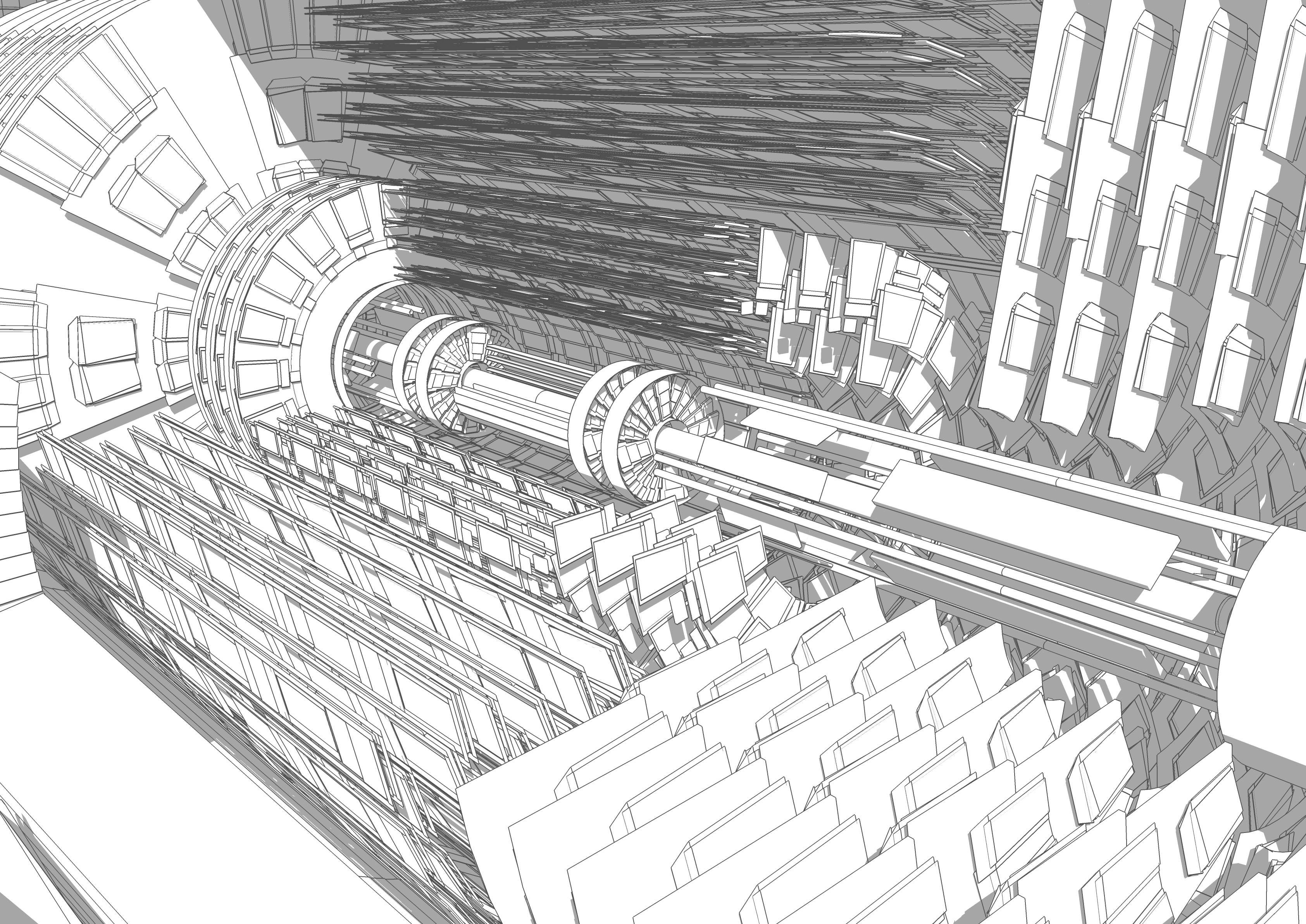}}
\caption{A detailed image of the CMS tracker.} \label{fig:zooom}
\end{figure*}

The level of detail in which to render the detector can be adjusted by
choosing subcomponents to include in the model. For example, Figure
\ref{fig:eb-module} shows a module of the \textit{barrel
electromagnetic calorimeter} (EB), in which the geometry of each
lead-tungstate ($\textrm{PbWO}_4$) crystal is drawn. Figure
\ref{fig:zooom} includes detailed geometry of the innermost
subsystems, the \textit{silicon strip} and \textit{pixel trackers}.
This figure is part of the exhibition \textit{``ZOOOM''}~\cite{zooom},
displayed at \textit{Point 5}, a CERN site in Cessy France, where the
CMS detector is installed.

\begin{figure*}[!h]
\centering
\includegraphics[width=9cm]{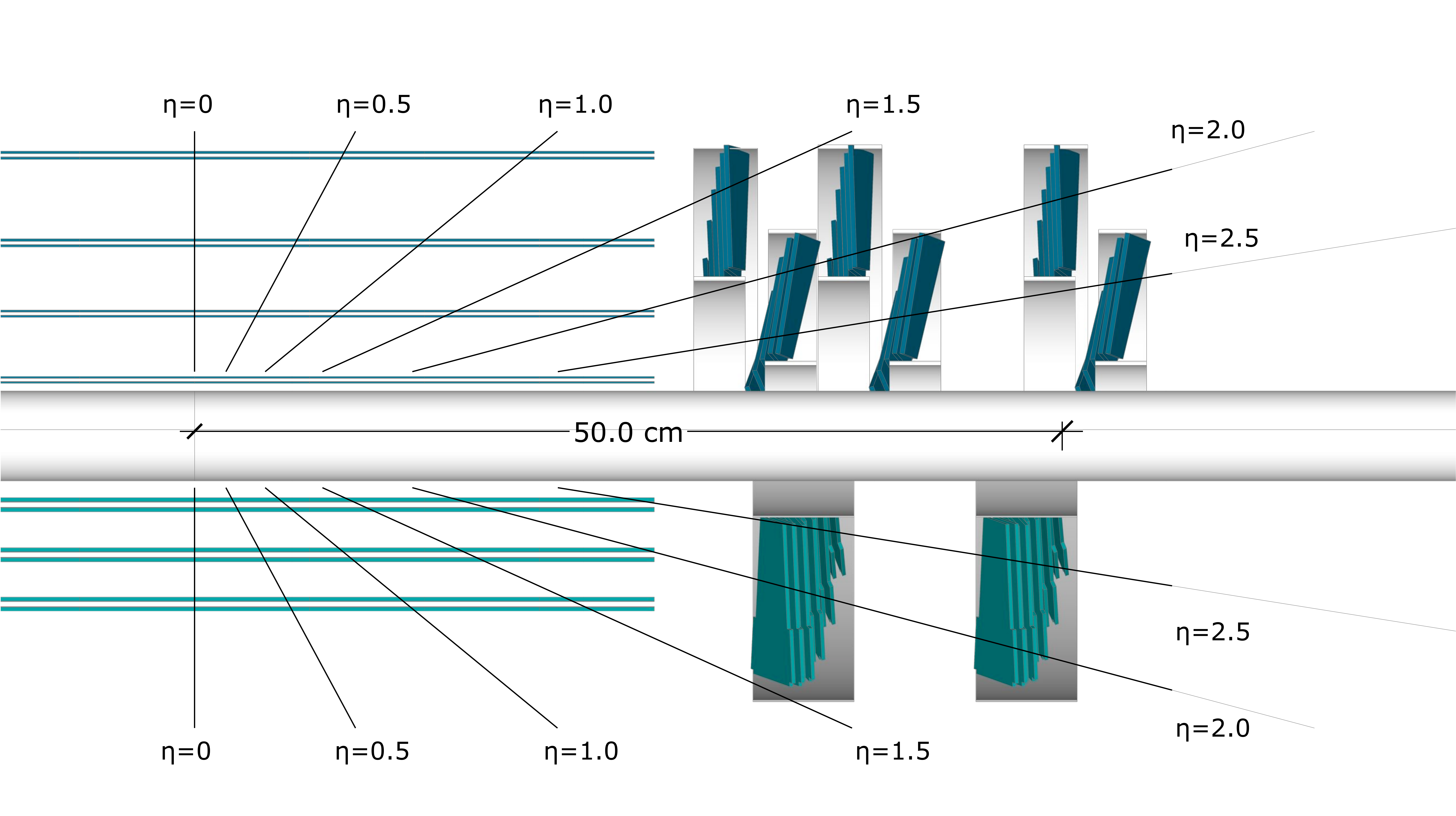}
\includegraphics[width=6cm]{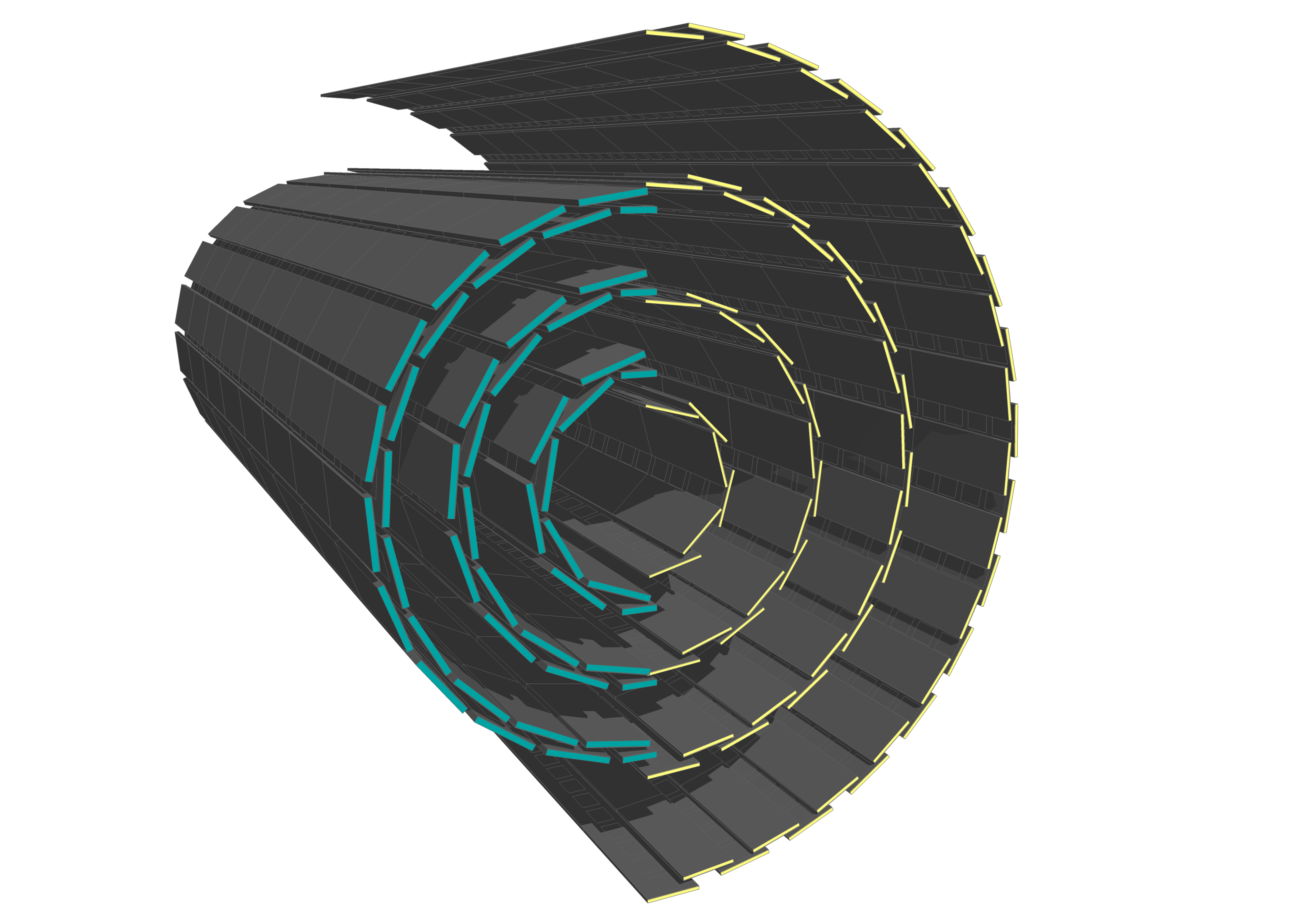}
\caption{\small Models of the CMS pixel tracker consisting of half initial
and half upgrade geometries.}
\label{fig:pixel-image}
\end{figure*}

The CMS detector will be continuously upgraded. The geometries for the
detector upgrades, which are also described in the CMS Detector
Description~\cite{cms-geom2020}, can be rendered in SketchUp as well.
In Figure \ref{fig:pixel-image}, one half of each model has the
initial geometry of the CMS pixel tracker and the other half has the
geometry for its \textit{Phase 1 upgrade}. These two images were used
to illustrate the difference between before and after the upgrade in
many public documents, including the \textit{CMS Technical Design
Report for the Pixel Detector Upgrade}~\cite{pixel-upgrade-tdr}.

\section{Event Rendering}

\begin{figure*}[!t]
\centering
\includegraphics[width=14.5cm]{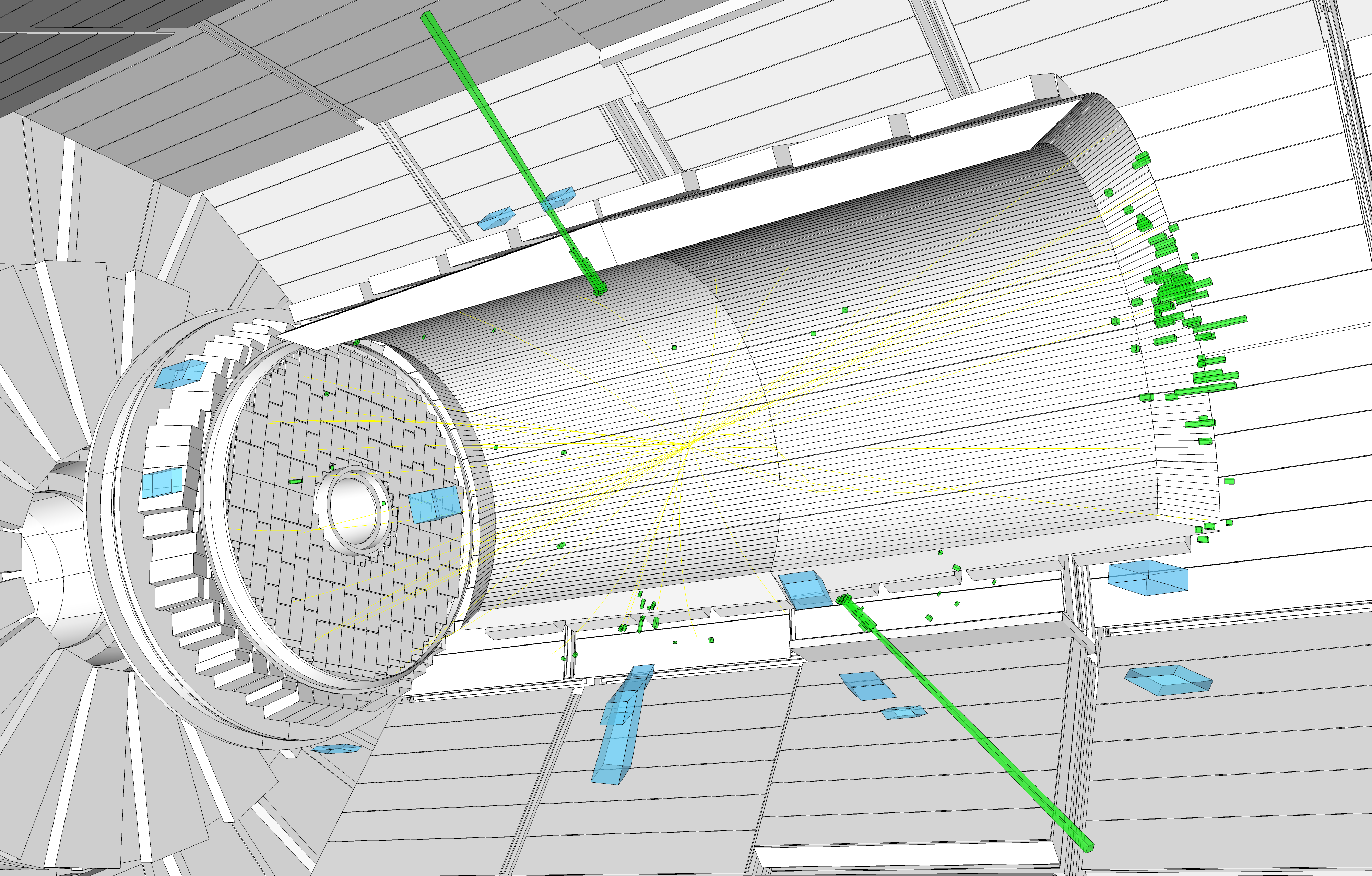}
\caption{\small A candidate Higgs boson decaying into two photons
(large green towers)} \label{fig:higgs-skp}
\end{figure*}

The event input is the \textit{ig} format~\cite{ispy-chep2012}
developed for the iSpy event display~\cite{ispy-website}. The content
includes information necessary for 2D and 3D rendering of event data
such as kinematical properties of reconstructed particles and
positions and energies of calorimetric deposits. The ig format is
simply a zip archive containing event files in JSON (JavaScript Object
Notation) format; a JavaScript object instance can be easily converted
to a Ruby hash. The SketchUp Ruby API can therefore easily be used to
render the events. In the larger project, a Ruby module named
\textit{ig2rb} parses an unzipped ig file and renders a subset of
objects. Tracks and electrons are rendered as \textit{cubic b\'{e}zier
splines}, energy deposits in the electromagnetic and hadronic
calorimeters as scaled rectangles with six faces, muons as
\textit{polylines}, and hit muon chambers as rectangular boxes. An
example image of a candidate Higgs boson decaying into two photons can
be seen in Figure~\ref{fig:higgs-skp}. Versions using other color sets
and styles can be found at the CERN Document Server~\cite{skp-cds}.

\section{Summary}

We have developed a set of Ruby scripts to create 3D computer models
of the CMS detector and CMS events in SketchUp via its Ruby API. These
scripts are available at its GitHub repository. The 3D models allow us
to produce high quality images and exportable 3D models of the CMS
detector and events, which are used in a variety of public
presentations. In addition, these models can be used to validate
geometries for CMS detector upgrades. Furthermore, several art
projects, exhibitions, outreach programs using these models are being
planned. We plan to continue to develop the scripts to extend the
range of supported reconstructed objects, detector subsystems, and
input formats and to improve the user interfaces for both interactive
and batch uses.

\newpage

\section*{Acknowledgments}
We thank Michael Case and Ianna Osborne for explaining to us the
technical implementation details of the CMS Detector Descriptions. We
have received valuable feedback and support from Erik Gottschalk,
Lucas Taylor, Michael Hoch, David Barney, Achintya Rao, Harry Cheung,
Teruki Kamon, Elizabeth Sexton-Kennedy, and other members of the CMS
collaboration. This work was partially supported by the U.S.
Department of Energy and National Science Foundation.


\end{document}